\documentclass[a4paper,conference]{IEEEtran}
\usepackage{amsmath,amsfonts,amssymb,amscd,mathrsfs,eucal}
\usepackage{graphicx}

\newtheorem{theorem}{Theorem}
\newtheorem{corollary}[theorem]{Corollary}

\newtheorem{definition}[theorem]{Definition}

\def\r{\rho}
\def\d{\delta}
\def\e{\epsilon}
\def\g{\gamma}
\def\G{\Gamma}

\ifCLASSINFOpdf
\else
\fi
\begin{document}
%
\vspace{-5mm}
\title{On construction and analysis of sparse random matrices and expander graphs with applications to compressed sensing.}

\author{\IEEEauthorblockN{Bubacarr Bah}
\IEEEauthorblockA{Laboratory for Information and Inference Systems\\
\'Ecole Polytechnique F\'ed\'erale de Lausanne\\
Lausanne, Switzerland\\
Email: bubacarr.bah@epfl.ch}
\and
\IEEEauthorblockN{Jared Tanner}
\IEEEauthorblockA{Mathematics Institute and Exeter College\\
University of Oxford\\
Oxford, United Kingdom\\
Email: tanner@maths.ox.ac.uk}}
\vskip -5mm


%


\maketitle

\begin{abstract}
We revisit the probabilistic construction of sparse random matrices where each column has a fixed number of nonzeros whose row indices are drawn uniformly at random. These matrices have a one-to-one correspondence with the adjacency matrices of lossless expander graphs.  We present tail bounds on the probability that the cardinality of the {\em set of neighbors} for these graphs will be less than the expected value. The bounds are derived through the analysis of collisions in unions of sets using a {\em dyadic splitting} technique. This analysis led to the derivation of better constants that allow for quantitative theorems on existence of lossless expander graphs and hence the sparse random matrices we consider and also quantitative compressed sensing sampling theorems when using sparse non mean-zero measurement matrices.
\end{abstract}


%
\IEEEpeerreviewmaketitle


\section{Introduction}\label{sec:intro}
Sparse matrices are particularly useful in applied and computational mathematics because of their low storage complexity and fast implementation as compared to dense matrices. Of late, significant progress has been made to incorporate sparse matrices in compressed sensing, with \cite{berinde2008combining,berinde2008sparse,jafarpour2009efficient,xu2007further} giving both theoretical performance guarantees and also exhibiting numerical results that shows sparse matrices coming from expander graphs can be as good sensing matrices as their dense counterparts. In fact, Blanchard and Tanner \cite{blanchard2012gpu}  recently demonstrated in a GPU implementation how well these type of matrices do compared to dense Gaussian and Discrete Cosine Transform matrices even with very small fixed number of nonzeros per column (as considered here).

In this manuscript we consider random sparse matrices that are adjacency matrices of lossless expander graphs. Expander graphs are highly connected graphs with very sparse adjacency matrices, a precise definition of a lossless expander graph is given in Definition \ref{def:expander}.
\begin{definition}
\label{def:expander}
$G \left(U,V,E\right)$ is a lossless $(k,d,\epsilon)$-expander if it is a bipartite graph with $|U| = N$ left vertices, $|V| = n$ right vertices and has a regular left degree $d$, such that any $X \subset U$ with $|X| \leq k$ has a set of neighbors $\Gamma(X) \subset V$ with $|\Gamma(X)| \geq \left(1 - \epsilon \right) d |X|$ neighbors.
\end{definition}

Note that these graphs are  {\em lossless} because $\e\ll1$, they are also referred to as {\em unbalanced expanders} in the literature because $n\ll N$ and a $(k,d,\e)$-lossless expander graph has an {\em expansion} of $\left(1 - \epsilon \right) d$. Such graphs have been well studied in theoretical computer science and mathematics and have many applications. Probabilistic constructions of such graphs using random left-regular bipartite graphs with optimal parameters exist but deterministic constructions only achieve sub-optimal parameters, see \cite{capalbo2002randomness} or \cite{hoory2006expander} for a more detailed survey.

Using a novel technique of {\em dyadic splitting of sets}, this work derives quantitative guarantees on the probabilistic construction of these graphs in the form of a bound on the tail probability of the size of the {\em set of neighbors}, $\Gamma(X)$ for a given $X \subset U$, of a randomly generated left-degree bipartite graph. Moreover, this tail bound proves a bound on the tail probability of the {\em expansion} of the graph, $|\G(X)|/|X|$. In addition, we derive the first phase transitions showing regions in parameter space that depicting when a left-regular bipartite graph with a given set of parameters is guaranteed to be a lossless expander with high probability. Similar results in terms of the adjacency matrices of these graphs is also presented. Another contribution of this work is the derivation of sampling theorems comparing performance guarantees for some of the algorithms proposed for compressed sensing using such sparse matrices as well as the more traditional $\ell_1$ minimization compressed sensing formulation. It also provides phase transitions of $\ell_1$ minimization performance guarantees for such sparse matrices compared to what $\ell_2$ restricted isometry constants ($\mathrm{RIC}_2$) analysis yields for Gaussian matrices.

\section{Tail Bound}\label{sec:tail}
Our main result is the presentation of formulae for the expected cardinality of the {\em set of neighbors} of $(k,d,\e)$-lossless expander graphs and the sparse non-mean zero matrices from these graphs. Based on this, we present a tail bound on the probability that this cardinality will be less than the expected value. We start by defining the class of matrices we consider and a key concept of a {\em set of neighbors} used in our derivation.
\begin{definition}
 \label{def:1set_expansion}
Let $A$ be an $n\times N$ matrix with $d$ nonzeros in each column. We refer to $A$ as a random a) sparse expander (SE) if every nonzero has value $1$ and b) sparse signed expander (SSE) if every nonzero has value from $\{-1,1\}$.
\end{definition}

The support set of the $d$ nonzeros per column of these matrices are drawn uniformly at random and independently for each column. An SE matrix is an adjacency matrix of $(k,d,\e)$-lossless expander graph while an SSE matrix have random sign patterns in the nonzeros of an adjacency matrix of a $(k,d,\e)$-lossless expander graph. If $A$ is either an SE or SSE it will have only $d$ nonzeros per column and since we fix $d\ll n$, $A$ is therefore extremely sparse. 

We formally define the {\em set of neighbors} in both graph theory and linear algebra notation to aid translation between the terminology of the two communities. Denote $A_S$ as a submatrix of $A$ composed of columns of $A$ indexed by the set $S$ with $|S|=s$.  
\begin{definition}
 \label{def:neighbours}
Consider a bipartite graph $G(U,V,E)$ where $E$ is the set of edges and $e_{ij}=(x_i,y_j)$ is the edge that connects vertex $x_i$ to vertex $y_j$. For a given subset of left vertices $S\subset U$ its set of neighbors $\Gamma(S) \subset V$ is defined as $\Gamma(S) := \{y_j|x_i\in S \mbox{ and } e_{ij}\in E\}$. In terms of the adjacency matrix, $A$, of $G(U,V,E)$ the set of neighbors of $A_S$ denoted by $A_s$, is the set of rows with at least one nonzero.
\end{definition}

Henceforth, we will only use the linear algebra notation $A_s$ which is equivalent to $\Gamma(S)$. Note that $\left|A_s\right|$ is a random variable depending on the draw of the set of columns, $S$, for each fixed $A$. Therefore, we can ask what is the probability that $\left|A_s\right|$ is not greater than $a_s$, in particular where $a_s$ is smaller than the expected value of $\left|A_s\right|$. This is the question that Theorem \ref{thm:prob_bound_1set_expansion} attempts to answers.

\begin{theorem}[Theorem 1.6, \cite{bah2012vanishingly}]
\label{thm:prob_bound_1set_expansion}
For fixed $s,n,N,d$ and $d\leq a_s <\infty$, let an $n\times N$ matrix, $A$ be drawn from either of the classes of matrices defined in Definition \ref{def:1set_expansion}, then
\begin{equation}
\label{eq:prob_bound_1set_expansion}
 \hbox{Prob}\left(\left|A_s\right| \leq a_s\right) < p_{max}(s,d) \cdot e^{\left[n\cdot\Psi\left(a_s,\ldots,a_1\right)\right]}
\end{equation}
where $p_{max}(s,d) = \frac{2}{25\sqrt{2\pi s^3d^3}}$, and for random variables $a_s, \ldots, a_2$ and $a_1:=d$,  $\Psi\left(a_s,\ldots,a_1\right)$ is given by
\begin{multline*}
\frac{1}{n} \bigg{[}3s\log\left(5d \right) + \sum_{i=1}^{\lceil s/2\rceil} \frac{s}{2i}\left(\left(n-a_i\right) \cdot \hbox{H}\left(\frac{a_{2i}-a_i}{n-a_i}\right) \right. \\ \left. + a_i\cdot \hbox{H}\left(\frac{a_{2i}-a_i}{a_i}\right) - n\cdot \hbox{H}\left(\frac{a_i}{n}\right) \right) \bigg{]},
\end{multline*}
where $\hbox{H}(\cdot)$ is the Shannon entropy function of base $e$ logarithm. Consequently: 
\begin{enumerate}
\item if no restriction is imposed on $a_s$ then the $a_i$ for $i>1$ take on the expected values of $\left|A_s\right|$, which are given by $ \hat{a}_{2i} = \hat{a}_{i}\left(2 - \frac{\hat{a}_{i}}{n}\right)$ for $i=1,2,4,\ldots,\lceil s/2\rceil$; 
\item else if $a_{s}$ is restricted to be less than $\hat{a}_{s}$, then the $a_i$ for $i>1$ are the unique solutions to the following polynomial system $ a_{2i}^3 - 2a_ia_{2i}^2 + 2a_i^2a_{2i} - a_i^2a_{4i} = 0$ for $i = 1, 2,\ldots,\lceil s/4\rceil$ with $a_{2i}\ge a_i$ for each $i$.
\end{enumerate}
\end{theorem}

\begin{figure}[h]
\centering
\includegraphics[width=0.4\textwidth]{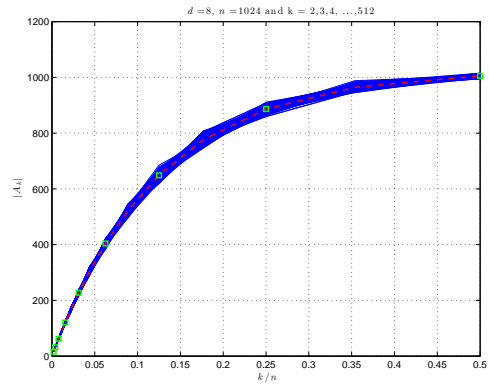}
\caption{For fixed $d=8$ and $n=2^{10}$, over $500$ realizations, plots (in blue) of the cardinalities of the index sets of nonzeros in a given number of set sizes, $k$. The dotted red curve is mean of the simulations and the green squares are the $\hat{a}_k$.}
\label{fig:cardinalities1}
\end{figure}

Theorem \ref{thm:prob_bound_1set_expansion} gives a bound on the probability that the cardinality of a union of $k$ sets each with $d$ elements is less than $a_k$. Figure \ref{fig:cardinalities1} shows plots of values of $\left|A_k\right|$ (size of set of neighbors) for different $k$ taken over 500 realizations (in blue), superimposed on these plots is the empirical mean values of $\left|A_k\right|$ over the 500 runs (in red) and the $\hat{a}_k$ in green.
\vspace{-2mm}
\vspace{-2mm}

\begin{figure}[h]
\centering
\includegraphics[width=0.4\textwidth]{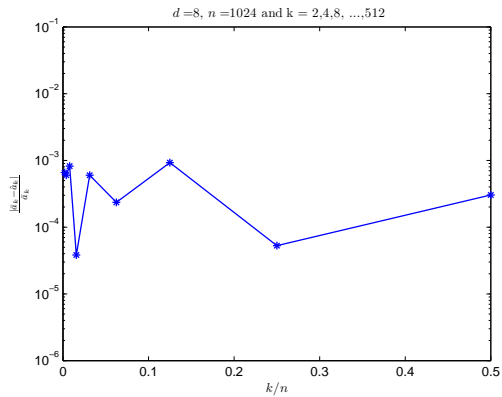}
\caption{For fixed $d=8$ and $n=2^{10}$, over $500$ realizations, plots of the relative error between the mean values of $a_k$ (referred to as $\bar{a}_k$) and the $\hat{a}_k$.}
\label{fig:cardinalities2}
\end{figure}

Furthermore, simulations illustrate that the $\hat{a}_k$ are the expected values of the cardinalities of the union of $k$ sets, $\left|A_k\right|$, as shown in Figure \ref{fig:cardinalities2}, where we show the relative error between $\hat{a}_k$ and the empirical mean values of the $\left|A_k\right|$, denoted by $\bar{a}_k$, realized over $500$ runs, to be less than $10^{-3}$.
\vspace{-2mm}
\vspace{-2mm}

\section{Sampling Theorems}\label{sec:sample}

We now use Theorem \ref{thm:prob_bound_1set_expansion} with the $\ell_1$-norm restricted isometry property (RIP-1), introduced by Berinde et. al. in \cite{berinde2008combining}, to deduce the corollaries that follow which are about the probabilistic construction of expander graphs, the matrices we consider, and sampling theorems of some selected compressed sensing algorithms. Firstly, using only the expansion property of these graphs we can draw the following corollary from Theorem \ref{thm:prob_bound_1set_expansion}.
\begin{corollary}
\label{cor:prob_bound_1set_expander}
For fixed $s,n,N,d$ and $0<\e<1/2$, let an $n\times N$ matrix, $A$ be drawn from the class of matrices defined in Definition \ref{def:1set_expansion}, then
\begin{equation*}
\label{eq:prob_bound_1set_expander}
 \hbox{Prob}\left(\mathop{\|A_Sx\|_1} \leq (1-2\e)d\|x\|_1\right) < p_{max}(s,d) \cdot e^{\left[n\cdot\Psi\left(s,d,\e\right)\right]},
\end{equation*}
where $\Psi\left(s,d,\e\right) = \Psi\left(a_s,\ldots,a_1\right)$ with $a_s = (1-\e)ds$.
\end{corollary}

Theorem \ref{thm:prob_bound_1set_expansion} and Corollary \ref{cor:prob_bound_1set_expander} allow us to calculate $s,n,N,d,\e$ where
the probability of the probabilistic constructions in Definition \ref{def:1set_expansion} not being a $(s,d,\e)$-lossless expander is exponentially small. Using Corollary \ref{cor:prob_bound_1set_expander} and the RIP-1 results in \cite{berinde2008combining} we derived a bound for the probability that a random draw of a matrix with $d ~1$s or $\pm 1$s in each column fails to satisfy the lower bound of the RIP-1 constant ($\mathrm{RIC}_1$) and hence fails to come from the class of matrices given in Definition \ref{def:1set_expansion}, for details see \cite{bah2012vanishingly}. From this bound we deduce the following corollary which is a sampling theorem on the existence of lossless expander graphs.

\begin{corollary}
\label{cor:prob_expander_existence2}
Consider $0<\e<1/2$ and $d$ fixed.   If $A$ is drawn from the class of matrices in Definition \ref{def:1set_expansion} and any $k$-sparse $x$ with $(k,n,N) \rightarrow \infty$ while $k/n \rightarrow \r \in (0,1)$ and $n/N \rightarrow \d \in (0,1)$ then for $\r < (1-\g)\r^{exp}(\d;d,\e)$ and $\g>0$
\begin{equation}
\label{eq:prob_expander_existence2}
\hbox{Prob}\left(\|Ax\|_1 \geq (1-2\e)d\|x\|_1\right) \rightarrow 1
\end{equation}
exponentially in $n$, where $\r^{exp}(\d;d,\e)$ is the largest limiting value of $k/n$ for which $\hbox{H}\left(\frac{k}{N}\right) + \frac{n}{N}\Psi\left(k,d,\e\right) = 0.$
\end{corollary}

For each fixed $0<\e<1/2$ and each fixed $d$, $\r^{exp}(\d;d,\e)$ in Corollary \ref{cor:prob_expander_existence2} is a function of $\d$ and a phase transition function in the $(\d,\r)$ plane. Below the curve of $\r^{exp}(\d;d,\e)$ the probability in \eqref{eq:prob_expander_existence2} goes to one exponentially in $n$ as the problem size grows. That is if $A$ is drawn at random with $d ~1$s or $d ~\pm 1$s in each column and having parameters $(k,n,N)$ that fall below the curve of $\r^{exp}(\d;d,\e)$ then we say it is from the class of matrices in Definition \ref{def:1set_expansion} with probability approaching one exponentially in $n$. In terms of $|\Gamma(X)|$ for $X\subset U$ and $|X|\leq k$, Corollary \ref{cor:prob_expander_existence2} say that the probability $|\Gamma(X)| \geq (1-\e) dk$ goes to one exponentially in $n$ if the parameters of our graph lies in the region below $\r^{exp}(\d;d,\e)$. This implies that if we draw a random bipartite graphs that has parameters in the region below the curve of $\r^{exp}(\d;d,\e)$ then with probability approaching one exponentially in $n$ that graph is a $(k,d,\e)$-lossless expander.
\begin{figure}[h]
\centering
\includegraphics[width=0.4\textwidth]{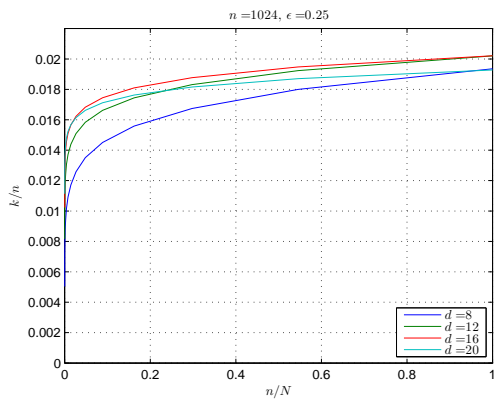}
\caption{Phase transition plots of $\r^{exp}(\d;d,\e)$ for fixed $\e=1/6$ and $n = 2^{10}$ with $d$ varied.}
\label{fig:phase_transition1}
\end{figure}

Figure \ref{fig:phase_transition1} shows a plot of what $\r^{exp}(\d;d,\e)$ converge to for different values of $d$ with $\e$ and $n$ fixed. It is interesting to note how increasing $d$ increases the phase transition up to a point then it decreases the phase transition. Essentially beyond $d=16$ there is inconsequential gain in increasing $d$. This vindicates the use of small $d$ in most of the numerical simulations involving the class of matrices considered here. Note the vanishing sparsity as the problem size $(k,n,N)$ grows while $d$ is fixed to a small value of $8$.

\begin{figure}[h]
\centering
\includegraphics[width=0.4\textwidth]{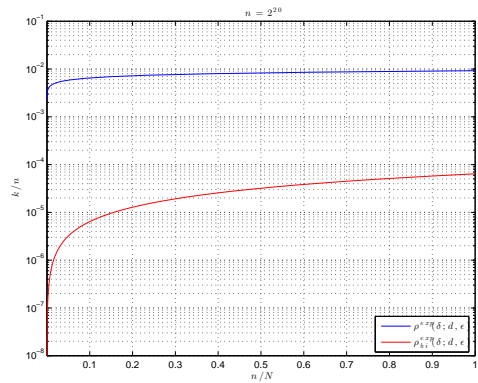}
\caption{A comparison of $\r^{exp}$ in Corollary \ref{cor:prob_expander_existence2} to $\r_{bi}^{exp}$ derived from the alternative construction proven in \cite{berinde2009advances}.}
\label{fig:phase_transition2}
\end{figure}

Corollary \ref{cor:prob_expander_existence2} can also be arrived at based on similar probabilistic constructions of expander graphs first proven by Pinsker in \cite{pinsker1973complexity} with more recent proofs in \cite{berinde2009advances,capalbo2002randomness}. To put our results in perspective, we compare them to the phase transitions derived from the constants from the construction in \cite{berinde2009advances}, shown in Figure \ref{fig:phase_transition2}.

Furthermore, for moderate values of $\e$ this allows us to make quantitative sampling theorems for some compressed sensing reconstruction algorithms. As usual in compressed sensing, in addition to $\ell_1$-minimization quite a few {\em combinatorial} greedy algorithms have been proposed for these sparse non-mean zero matrices. These algorithms iteratively locates and eliminate large (in magnitude) components of the vector, \cite{berinde2008combining}. They include Sequential Sparse Matching Pursuit (SSMP), see \cite{berinde2009sequential}; and Expander Recovery (ER), see \cite{jafarpour2009efficient}. Besides, theoretical guarantees have been given for $\ell_1$ recovery and some of the greedy algorithms including SSMP and ER. Base on these theoretical guarantees, we derived sampling theorems and present here phase transition curves which are plots of phase transition functions $\r^{alg}(\d;d,\e)$ of algorithms such that for $k/n \rightarrow \r < (1-\g)\r^{alg}(\d;d,\e), ~\g>0$, a given algorithm is guaranteed to recovery all $k$-sparse signals with overwhelming probability approaching one exponentially in $n$.

\begin{figure}[h]
\centering
\includegraphics[width=0.4\textwidth]{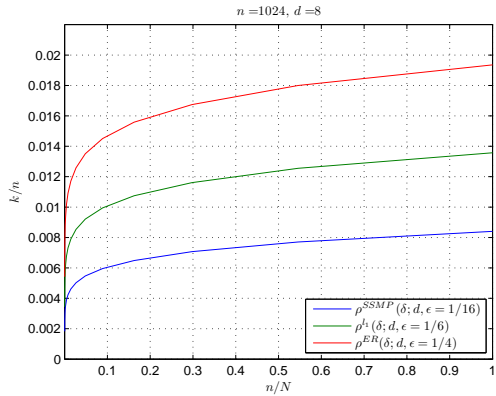}
\caption{Phase transition curves $\rho^{alg}\left(\delta;d,\e\right)$ computed over finite values of $\delta \in (0,1)$ with $d$ fixed and the different $\epsilon$ values for each algorithm - 1/4, 1/6 and 1/16 for ER, $\ell_1$ and SSMP respectively.}
\label{fig:pt_alg_compare1}
\end{figure}

\begin{figure}[h]
\centering
\includegraphics[width=0.4\textwidth]{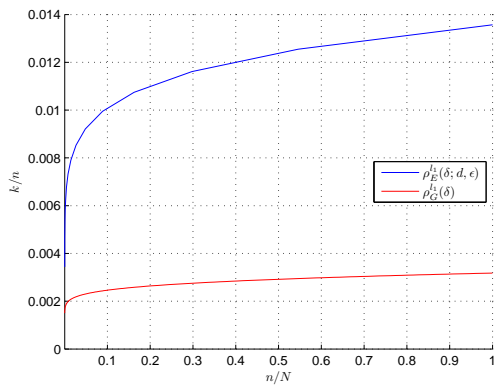}
\caption{Phase transition plots of $\ell_1$, $\rho^{\ell_1}_G\left(\delta\right)$, for Gaussian matrices derived using $\mathrm{RIC}_2$ and $\rho^{\ell_1}_E\left(\delta;d,\e\right)$ for adjacency matrices of expander graphs with $n=1024$, $d=8$, and $\e=1/6$.}
\label{fig:pt_alg_compare2}
\end{figure}

Figure \ref{fig:pt_alg_compare1} compares the phase transition of thee above mentioned algorithms. Remarkably, for ER recovery is guaranteed for a larger portion of the $(\delta,\rho)$ plane than is guaranteed by the theory for $\ell_1$-minimization using sparse matrices; however,  $\ell_1$-minimization has a larger recovery region than does SSMP. Figure \ref{fig:pt_alg_compare2} shows a comparison of the phase transition of $\ell_1$-minimization as presented by Blanchard et. al. in \cite{blanchard2011phase} for dense Gaussian matrices based on $\mathrm{RIC}_2$ analysis and the phase transition we derived here for the sparse binary matrices coming from lossless expander based on $\mathrm{RIC}_1$ analysis. This shows a significant difference between the two with sparse matrices having better performance guarantees.
However, these improved recovery guarantees are likely more due to the closer match of the method of analysis than to the efficacy of sparse matrices over dense matrices. 

\section{Sketch of Main Proof}\label{sec:proof}
Due to space constraints the details of the proofs are skipped and the interested reader is referred to \cite{bah2012vanishingly}. It is however important to briefly describe the key innovations in the derivation of the main result, Theorem \ref{thm:prob_bound_1set_expansion}.

For one fixed set of columns of $A$, denoted $A_S$,  the probability in \eqref{eq:prob_bound_1set_expansion} can be understood as the cardinality of the unions of nonzeros in the columns.  Our analysis of this probability follows from a nested unions of subsets using a {\em dyadic splitting} technique. Given a starting set of columns we recursively split the number of columns from this set and the resulting sets into two sets of cardinality of the ceiling and floor of the cardinality of their union until a level when the cardinalities are at most two. Resulting from this type of splitting is a regular binary tree where the size of each child is either the ceiling or the floor of the size of it's parent set. The probability of interest becomes a product of the probabilities involving all the children from the dyadic splitting of $A_s$. The proof therefore reduces to upper bounding this product.

Furthermore, in the binary tree resulting from our dyadic splitting scheme the number of columns in the two children of a parent node is the ceiling and the floor of half of the number of columns of the parent node. At each level of the split the number of columns of the children of that level differ by one. The enumeration of these two quantities at each level of the splitting process is necessary in the computation of the bound in \eqref{eq:prob_bound_1set_expansion}. This led to another novel technical result in our derivation, i.e. {\em dyadic splitting lemma} (Lemma 2.5 in \cite{bah2012vanishingly}).

\section{Conclusions}\label{sec:conclusion}
This work derived bounds on the tail probability of the cardinality of the {\em set of neighbours} of expander graphs resulting into better order constants than the standard probabilistic construction. Using this bound and $\mathrm{RIC}_1$ analysis, we deduce sampling theorems for the existence of expander graphs and their adjacency matrices. The derivation of the tail bound used a novel technique of {\em dyadic set splitting}. We also compared quantitatively, performance guarantees of compressed sensing algorithms which show greater phase transitions for ER than $\ell_1$-minimization which in turn is greater than SSMP.  A comparison of  $\ell_1$-minimization for dense and sparse matrices shows a higher phase transition for sparse matrices.

\end{document}